\documentstyle[12pt]{article}

\newcommand{\nc}{\newcommand}
\nc{\beq}{\begin{equation}}
\nc{\eeq}{\end{equation}}
\nc{\beqa}{\begin{eqnarray}}
\nc{\eeqa}{\end{eqnarray}}

{\nc{\lsim}{\mbox{\raisebox{-.6ex}{~$\stackrel{<}{\sim}$~}}}
{\nc{\gsim}{\mbox{\raisebox{-.6ex}{~$\stackrel{>}{\sim}$~}}}

\raggedbottom 
\title{\begin{flushright}
{\normalsize NUC-MINN-96/6-T\\
\normalsize NBI-96-24\\
May 1996\\}
\end{flushright}
\vspace*{0.3in}
{\bf DYNAMICAL FORMATION OF DISORIENTED CHIRAL CONDENSATES}}
\author{{\bf Joseph I. Kapusta}\\
 {\it School of Physics and Astronomy}\\
 {\it University of Minnesota}\\ \vspace*{0.2in} {\it Minneapolis, MN 55455}\\
{\bf Axel P. Vischer}\\
 {\it Niels Bohr Institute}\\
 {\it DK-2100 Copenhagen \O}\\
 {\it Denmark}}

\date{}

\parindent=20pt

\begin{document}

\maketitle

\begin{center}
Abstract
\end{center}

\noindent
We study the dynamical formation of disoriented chiral condensates
in very high energy nucleus-nucleus collisions using Bjorken
hydrodynamics and relativistic nucleation theory.  It is the
dynamics of the first order confinement phase transition
which controls the evolution of the system.  Every bubble or
fluctuation of the new, hadronic, phase obtains its own chiral condensate
with a probability determined by the Boltzmann weight of the finite
temperature effective potential of the linear sigma model.
We evaluate domain size and chiral angle distributions, which
can be used as initial conditions for the solution of semiclassical
field equations.

\newpage

\section{Introduction}

It is expected that quark--gluon plasma will be created in high energy
heavy ion collisions at the Relativisitic Heavy Ion Collider (RHIC)
at Brookhaven National Laboratory and at the Large Hadron Collider (LHC)
at CERN.  The experiments should allow us to study the chiral
and/or confinement phase transition/crossover.  One possible consequence
of plasma formation and disassembly is the creation of misaligned chiral
domains (disoriented chiral condensates or DCC) which are regions of
space--time where the average value of the chiral field does not point
in the same direction as the surrounding true vaccuum.  For
a brief review see \cite{gavin1}.

This possibility was really emphasized and studied by Rajagopal and
Wilczek \cite{wr1} in the context of heavy ion collisions within
a version of the linear $\sigma$--model.  It was found that if the
system evolves close to thermal equilibrium the average domain size
is of order $1/T_{\rm c}$, the inverse of the phase transition temperature.
The resulting domain sizes are therefore too small to have
many observable consequences; enhanced baryon--antibaryon production
may, however, be one of them \cite{Ajit}.  Subsequent studies
of DCC concentrated on nonequilibrium scenarios like quenching
\cite{quench} where the system is suddenly relaxed from an initial
thermal state above $T_{\rm c}$ to zero temperature.  The resulting
field configuration is unstable and decays on time scales of order
$1/m_{\sigma}$.  Again one obtains domain sizes which are too
small to have many experimental consequences.  Finally, in an
annealing scenario \cite{anneal}, medium modifications of
the $\sigma$ were taken into account, allowing for a reduction of
$m_{\sigma}$ and therefore an increase of domain size of
up to 3 fm.

All aforementioned approaches assume that the phase transition is, or
at least close to, second order.  In this paper we would like to
investigate the consequences of a first order phase transition.
The disoriented chiral domains are created via statistical fluctuations.
We assume that a relativistic heavy ion collision results in the
formation of an extended volume of quark gluon plasma which
transforms itself into a hadronic resonance gas through homogeneous
nucleation of hadronic bubbles \cite{k1}--\cite{k3}.  Bubbles of the new
hadronic phase are created by statistical fluctuations which grow
due to their lower free energy.  The phase transition is completed
when the entire plasma has been transformed into the hadronic resonance
gas.  This dynamics implements the confinement characteristics of
the phase transition.  The dynamical information is contained in the
free energy difference between the plasma phase and the resonance gas
phase and in their transport properties.

We assume that the dynamical evolution will be dominated by this
color confinement at temperatures close to the critical temperature
$T_{\rm c}$.  In principle we could imagine that every hadronic bubble
nucleated in this way starts out with the chiral field pointing
in a random direction in internal symmetry space.  The bubbles grow
with time until the whole plasma phase is transformed into the
hadronic phase, resulting in a distribution of domains with different
sizes and different chiral angles.  For definiteness we use the linear
$\sigma$--model to evalutate the part of the hadronic free energy
dependent on the chiral field.  The effect of this contribution on
the dynamical evolution is perturbative.

The domains formed in this approach will be larger or
of order $R_{\rm c}$, the critical size radius to nucleate an
hadronic bubble.  This is a dynamical quantity which diverges at
the critical temperature and approaches the hadronic length scale
$1/T_{\rm c}$ at small temperatures.  We calculate a domain size
and chiral angle distribution.  This distribution may be taken as
the initial condition for solving the semiclassical equations of
motion for the subsequent time evolution of the system~\cite{nordic}.

Our approach will neglect the possibility that a bubble nucleated
with a very large radius might contain several chiral domains.
This is not a serious concern since the average nucleation bubble
is not so large.  Furthermore, the resulting domain size and angle
distribution should also depend on the rapidity.  For
given rapidity the nucleated bubbles will obtain a certain
eccentricity in the beam direction. In
this first study we neglect this complication.

In section 2 we start out by outlining the implementation of the
dynamical evolution of the system.  The free energies we are using
in this section are at zero chiral angle.  In section 3 we evaluate
the perturbative chiral contribution to the hadronic free energy
at nonzero chiral angle and show how to obtain the domain size
and angle distribution.  In section 5 we discuss and summarize our results.

\section{Dynamics of the Confinement Phase Transition}

In this section we briefly review the dynamics of the hot matter
as it passes through a first order confinement phase transition.
The overall picture is that quark--gluon plasma is formed at
high temperature and subsequently expands and cools.  The plasma
must cool below the critical temperature before bubbles of the
hadronic phase can be nucleated.  When enough bubbles have been
nucleated their growth causes a reheating, turning off further
bubble nucleation.  Eventually all matter is converted to the
hadronic form, and cooling begins again.  Eventually the hadrons
lose thermal contact with each other and they free--stream to the
detectors.  It is within this scenario that we consider the
formation of DCC.  The works \cite{k1}--\cite{k3} should be consulted
for more details about this scenario.

\subsection{Review of nucleation dynamics}

The rate for the nucleation of the hadron phase out of the plasma phase
can be written as
\begin{equation}
I = I_0\,e^{-\Delta F_*/T},
\end{equation}
where $\Delta F_*$ is the change in the free energy of the system with
the formation of a critical size hadronic bubble and $I_0$ is the
prefactor with dimensions of inverse volume inverse time.
In general, statistical fluctuations at $T < T_{\rm c}$ will produce
bubbles with associated free energy
\begin{equation}
\Delta F = \frac{4\pi}{3}[p_{\rm q}(T)-p_{\rm h}(T)]R^3+4\pi R^2 \sigma.
\end{equation}
Here $p$ is the pressure of the quark or hadron phase at temperature $T$,
and $\sigma$ is the surface free energy of the quark-gluon/hadron interface.
Since $p_{\rm q} < p_{\rm h}$ it follows that there is a bubble of critical
radius
\begin{equation}
R_*(T) = {2\sigma}/[{p_{\rm h}(T)-p_{\rm q}(T)}].
\end{equation}
Smaller bubbles tend to shrink because the surface energy is too large
relative to volume energy, and larger bubbles tend to grow.  The free
energy of the critical size bubble is therefore
\begin{equation}
\Delta F_* = \frac{4}{3}\pi \sigma R^2_*.
\end{equation}
The prefactor has been computed in a coarse-grained
effective field theory approximation to QCD to be
\begin{equation}
I_0 = \frac{16}{3\pi}\left( \frac{\sigma}{3T} \right) ^{3/2}
\frac{\sigma \eta_{\rm q} R_*}{\xi_{\rm q}^4 (\Delta w)^2},
\end{equation}
where $\eta_{\rm q}$ is the shear viscosity in the plasma phase, $\xi_{\rm q}$
is a correlation length in the plasma phase, and $\Delta w$ is the
difference in the enthalpy densities of the two phases.
At the critical temperature, $R_* \rightarrow \infty$, and the rate vanishes.

Given the nucleation rate one would like to know the (volume) fraction of
space $h(t)$
which has been converted from QCD plasma to hadronic gas at the
proper time $t$, which is the time as measured in the local comoving frame
of an expanding system.  This requires a kinetic equation which uses
$I$ as an input.
If the system cools to $T_{\rm c}$
at time $t_{\rm c}$ then at some later time $t$ the fraction
of space which has been converted to hadronic gas is
\begin{equation}
h(t) = \int_{t_{\rm c}}^t dt' I(T(t')) [1-h(t')] V_{\rm bub}(t',t).
\label{dyn.1}
\end{equation}
Here $V_{\rm bub}(t',t)$ is the volume of a bubble at time $t$ which
had been nucleated at the earlier time $t'$.

Once formed, bubbles will grow, as it is favorable from the point
of view of free energy.  We assume the growth law
\begin{equation}
v(T) = v_0 [1-T/T_{\rm c}]^{3/2},
\end{equation}
where $v_0$ is a model-dependent constant.
The simple illustrative model for bubble growth is
\begin{equation}
V_{\rm bub}(t',t) = \frac{4\pi}{3}\left( R_*(T(t')) + \int_{t'}^t dt''
v(T(t'')) \right)^3.
\end{equation}
This expression assumes that the interface between the inside and
outside of the bubble is created at rest in the local comoving
frame.

A dynamical equation is needed to describe how the system expands.
We will use Bjorken's longitudinal scaling hydrodynamics \cite
{bj}.  The derivative of the energy density is related to the enthalpy
density as
\begin{equation}
\frac{de}{dt} \, = \, -\frac{w}{t}.
\end{equation}
This assumes kinetic equilibrium among the particles but not phase
equilibrium.  It is a statement of energy conservation.  The energy
density is
\begin{equation}
e(T) = h(t)e_{\rm h}(T)+[1-h(t)]e_{\rm q}(T),
\end{equation}
where $e_{\rm h}(T)$ and $e_{\rm q}(T)$ are the energy densities in the two
phases at temperature $T$, and similarly for $w$.

We model the plasma phase by a gas of gluons and massless quarks of
either two or three flavors with a bag constant $B$ to simulate
confinement.  The pressure is
\begin{equation}
p_{\rm q} = g_{\rm q} \frac{\pi^2}{90}T^4 - B \, ,
\end{equation}
where $g_{\rm q}$ is chosen appropriately.
We model the pressure in the hadronic phase according to the formula
\begin{equation}
p_{\rm h} = g_{\rm h} \frac{\pi^2}{90}T^4
\end{equation}
where $g_{\rm h}$ is an effective number of degrees of freedom relevant
for the temperature range of interest, namely, $130 < T < 170$ MeV.
Including noninteracting pions alone, for example, gives
$g_{\rm h} \approx 3$.  Here we will consider two parameter sets.
Set A includes only $u$ and $d$ quarks.  In the hadron phase we
include the particles $\pi, \eta, \omega, \rho$.  In the aforementioned
temperature range $g_{\rm h} \approx 4.6$ \cite{D1}.  Set B includes
$u, d$ and $s$ quarks.  In the hadron phase we include all states
in the Particle Data Book with a repulsive interaction among them.
This equation of state was studied in \cite{Olive}.  A good parametrization
is obtained with $g_{\rm h} \approx 7.5$.  The critical temperature
is obtained by equating the pressures of the two phase.  We adjust
the bag constant so that $T_{\rm c} = 160$ MeV.  In set A one gets
$B^{1/4} = 220$ MeV and in set B one gets 232 MeV.
The other parameters are as in \cite{k2}, namely: $\sigma = 50$
MeV/fm$^2$, $\xi_{\rm q} = 0.7$ fm, $\eta_{\rm q} = 14.4 T^3$ and $v_0 = 3$.

\subsection{Numerical results}

The evolution equations may be solved as in \cite{k2}.  Figure 1
shows the temperature as a function of time.  It is assumed that
the system cools to $T_{\rm c}$ at a local comoving time of 3 fm/c.
The system must supercool to about 0.95$T_{\rm c}$ before noticeable
nucleation begins.  The system continues to expand and cool due
to inertia, but is slowed down by bubble nucleation and growth.
The temperature reaches a local minimum at around 0.80$T_{\rm c}$,
at which point the release of latent heat is sufficient to begin
reheating.  As the temperature goes up, the nucleation rate
decreases, and basically shuts off at about 0.95$T_{\rm c}$.  Thereafter
there is essentially no new bubble creation; the transition
completes due to the growth of existing bubbles.  When the
fraction of space $h$ occupied by hadrons reaches 100\% the
transition is complete and the temperature will again fall.
This is not shown on the figure.  It is the state of the system
at this moment which will determine the initial conditions
for solving the equations of motion for the chiral fields.

Figure 2 shows the volume of a hadronic bubble at the end of
the phase transition as a function of the time at which it was
nucleated.  The volume of a critical size bubble is a monotonically
decreasing function of temperature.  This means that bubbles which
were nucleated early or late in the supercooling phase should
have the largest volume, as is seen to be the case.  The smallest
bubbles should be those which were nucleated near the minimum of the
the temperature curve, which is also seen to be the case.  The minimum
of the volume curve is offset to somewhat later times than the
minimum of the temperature curve due to the subsequent growth of
bubbles.

Figure 3 shows the density of bubbles as a function of their
volume at the end of the transition.  This is obtained by keeping
track of how many bubbles were nucleated between times $t$ and
$t + dt$ and then following their growth to the end of the
transition at time $t_{\rm f}$.  The density distribution is
\begin{equation}
\frac{dn}{dV} = \int_{t_{\rm c}}^{t_{\rm f}} \frac{dt\,t}{t_{\rm f}}
I(t) [1-h(t)]
\delta\left( V - V_{\rm bub}(t,t_{\rm f}) \right) \, ,
\end{equation}
where the factor $t/t_{\rm f}$ takes into account the fact that the
density decreases due to the expansion of the system.  In general
there will be two times contributing to the same final bubble
volume, as already seen in figure 2.  The distribution plotted
in figure 3 is made dimensionless by multiplying by the minimum
volume squared and is plotted against the linear size of the
bubble to better display the results.  For set A the minimum
final volume is 22 fm$^3$ while for set B it is 10 fm$^3$.
It should be remarked that at the end of the transition the
bubbles will not be spherical due to the longitudinal, Bjorken,
expansion along the beam axis.  They will be elongated along
the beam direction with a typical eccentricity of $t_{\rm f}/t_{\rm sc}
\approx 3 - 4$ where $t_{\rm sc}$ is the time at which the temperature
has dropped to its minimum value along the supercooling curve.

Some important numbers from the supercooling and reheating cycle
are listed in Table 1.  For each paramter set we list:
the time $t_{\rm sc}$ and temperature $T_{\rm sc}$ when supercooling ends
and reheating begins; the completion time $t_{\rm f}$ and temperature
$T_{\rm f}$ of the phase transition; the minimum volume of a bubble
$V_{\rm min}$ at the end of the transition and the time $t_{\rm min}$
at which it was nucleated.

The distribution $dn/dV$ at the end of the transition diverges
at its lower limit like $1/\sqrt{1-V_{\rm min}/V}$.  This is a
rather weak, integrable, singularity which may, however, have
important consequences for the observation of DCC in heavy ion
experiments.

One remarkable feature of this dynamics is that the contribution
to large bubbles comes mainly from early times, $t < t_{\rm sc}$.
Bubbles which are nucleated near the bottom of the temperature curve
do not have sufficient time to grow very much before reheating
causes the growth velocity to decrease substantially.  Bubbles nucleated
at later times have a larger (critical) volume to begin with, but
their growth is also suppressed by a decreasing growth velocity.
In addition, their rate of production is suppressed relative to
early times due to the factor $1-h$ in eq. (13); there is insufficient
space available between existing bubbles to nucleate new ones.
The distribution for volumes larger than about 100 fm$^3$ is almost
completely determined by the initial stages of the phase transition.
Even if we would shut--off nucleation by hand at time $t_{\rm sc}$
we would essentially obtain the same size distribution
at large volumes. The reheating phase of the transition does not
influence the size distribution at large volumes!

The behavior of $dn/dV$ at large volumes is well fitted to the
form $\exp[-a\, (V/V_{\rm min})^{2/3}]$.  The values of the
coefficient $a$ are given in Table 1.  The exponential dependence
on the surface area is straightforward.  Large bubbles at the
end of the transition arise from large bubbles nucleated early on.
The probability to nucleate one is proportional to the rate
obtained from eqs. (1) and (4), which is exponential in the
surface area of a critical bubble.  Allowing for growth due
to longitudinal expansion we can estimate roughly that
\beqa
a = \left(\frac{4 \pi}{3}\right)^{1/3}\,\frac{\sigma}{T_{\rm c}}
\,\left(\frac{t_0}{t_{\rm f}}\right)^{2/3}\,
V_{\rm min}^{\frac{2}{3}}\,.
\eeqa
If we compare this result with the fitted values given in Table 1
we get agreement to better than a factor of 2.

\section{DCC Within a Nucleated Bubble}

In the previous section we discussed how a first order phase transition
from the deconfined quark--gluon phase to the confined hadronic phase
may occur via supercooling, nucleation, bubble growth, and reheating.
We did not allow for the possibility that a disoriented chiral condensate
could appear spontaneously with the nucleation of the bubble.
In this section we will.  For simplicity and definiteness we will
use the linear $\sigma$--model to implement approximate chiral symmetry.
The essence of this approach is that the free energy difference
$\Delta F(R)$ in eq. (2) contains all hadronic degrees of freedom,
including the pions and the $\sigma$--meson, but at zero chiral angle.
To describe now chiral condensates at high temperatures we use the ansatz
\beqa
\Delta F(R,\theta) =
\Delta F(R,0) + \Delta F_{\rm chiral}(R,\theta)\,,
\label{ansatz}
\eeqa
where $\Delta F_{\rm chiral}$ is a perturbative contribution to the
free energy difference between hadronic and quark--gluon phases
for nonzero chiral angle; it is normalized such that $\Delta F_{\rm chiral}
(R,0) = 0$.  For low temperatures the system will tend to orient itself
towards $\theta = 0$.  On the other hand, if the temperature is larger
than a certain chiral critical temperature $T_{\rm ch}$ all angles should
be approximately equally likely.  The aim of this section is to
construct $\Delta F_{\rm chiral}(R,\theta)$ and to show how to obtain,
perturbatively, a distribution in the chiral angle over an ensemble of
differently-sized disoriented chiral domains.

\subsection{The effective potential for chiral angles}

We will evaluate the finite temperature effective potential representing
the degrees of freedom of the chiral condensate by using the linear
sigma model.
\beqa
{\cal L} = \frac{1}{2}\, (\partial_{\mu} \sigma)^2 +
 \frac{1}{2}\, (\partial_{\mu} {\bf \pi})^2
-\frac{\lambda}{4}\,\left(\sigma^2+{\bf \pi}^2-c^2/\lambda\right)^2
 + H \sigma \,.
\label{lagrangian}
\eeqa
The parameters of the Lagrangian $\cal L$ are fixed at tree
level by imposing the existence of spontaneous symmetry breaking,
PCAC and the meson masses in vaccuum.  This leads to the condensates
$\langle \sigma \rangle = v$ and $\langle \pi_{\rm i} \rangle = 0$ where
\beqa
v(\lambda v^2 -c^2) - H &=& 0 \, , \nonumber \\
H&=&f_{\pi}\,m_{\pi}^2\,,\nonumber\\
m_{\pi}^2 &=& \lambda\,v^2-c^2\,,\nonumber\\
m_{\sigma}^2 &=& m_{\pi}^2+2\,\lambda\,v^2\,.
\label{conditions}
\eeqa
Numerical values used here are: $f_{\pi}=94.5$ MeV, $m_{\pi}=140$ MeV,
and $m_{\sigma} =1$ GeV.  The value of the $\sigma$--mass is chosen
in accordance with the analysis of reference \cite{sigma}.
The evaluation of the system of equations in (\ref{conditions})
yields $v=f_{\pi}$, $H=47.6$ MeV/fm$^2$, $\lambda=54.9$ and $c=686$ MeV.

The finite temperature effective potential for the linear sigma model
is discussed in many places; for example, in \cite{proxi}.
Since eq. (\ref{lagrangian}) has a remnant O(3) symmetry
in the isovector sector we can simplify this task by letting
the condensate point in the third direction of the isovector space.
To evaluate the thermal masses we expand the fields around an arbitrary
point as:
\beqa
\sigma(x) &=& v \,\cos\theta+\sigma'(x)\,,\nonumber\\
\pi_3(x)  &=& v\,\sin\theta+\pi_3'(x)\,.
\label{newfields}
\eeqa
We find
\begin{equation}
m_1^2=m_2^2 = m_3^2 =\lambda v^2-c^2\,,
\end{equation}
\begin{equation}
m_0^2=3\lambda v^2-c^2\,.
\end{equation}
In the high temperature limit of the one-loop approximation one
keeps terms that are of order $T^4$ and $m^2T^2$ only.  Then the
effective potential is:
\beq
V_{\rm eff}(T;v,\theta) = \frac{\lambda}{4}
\left[ v^4-\left( T_{\rm ch}^2-T^2 \right) v^2 \right]
- H\,v\,\cos\theta\,.
\label{veff}
\eeq
The chiral critical temperature is defined by $T_{\rm ch}=
{\sqrt{2c^2/\lambda}}$.  It is the temperature where the curvature term
in the effective potential vanishes.  When chiral symmetry is exact,
$m_{\pi}=0$, and there is a second order phase transition associated
with chiral symmetry restoration.  For our choice of parameters
this temperature is $130.9$ MeV.  It is useful to recognize that
for a fixed chiral angle $\theta$ the effective potential is minimized
when $v$ satisfies the cubic equation
\beq
v^3 - \frac{1}{2}\left(T^2_{\rm ch} - T^2 \right)v
- \frac{H}{\lambda}\,\cos\theta = 0\,.
\label{vcl}
\eeq
We shall label solutions to this equation as $v_{\theta}(T)$.

To develop an intuitive understanding of the dynamics of the chiral
contribution we first plot, in figure 4, the value of the chiral
condensate at the global minimum of the effective potential
as a function of the temperature.  The global minimum is always
at zero chiral angle when chiral symmetry is dynamically broken by
nonzero pion mass.  When chiral symmetry is exact the potential
is independent of angle.  The dashed curve in the figure represents
the chiral limit $m_{\pi} = 0$ in which we recover a second order
chiral phase transition; the chiral critical temperature approaches
133.6 MeV when keeping the $\sigma$--mass and pion decay constant
fixed.  With the physical value of the pion mass the chiral condensate
never goes to zero; the minimum of the effective potential is always
away from the origin and chiral symmetry cannot be restored.  Nevertheless
the chiral condensate $v_0(T)$ becomes very small at high temperatures,
being about 7.9 MeV at the confinement temperature of 160 MeV.
For high temperatures it decreases like $2 /\lambda T^2$ according to
equation (\ref{vcl}).  The minimum of the effective potential itself
in (\ref{veff}) decreases like $2 H^2/\lambda T^2$.
We can now compare the magnitude of the pressure difference of the
quark--gluon and hadronic phases in eqs. (11)-(12) to the magnitude
of the effective potential for the chiral angle $\theta$.
For more than 2$\%$ supercooling the chiral contribution is perturbative.
There is essentially no nucleation for such small supercooling so
the chiral contribution is safely in the perturbative region.

In figure 5 we plot the isotherms of the function $v_{\theta}(T)$
in the $\pi_3$--$\sigma$ plane.  At low temperatures the isotherms
are approximately circles with radii $\leq f_{\pi}$.  The radius
shrinks and the circle becomes more distorted as the temperature increases.
For temperatures less than
\beq
T_{\rm saddle} = {\sqrt{T_{\rm ch}^2-
6\left(\frac{H}{2\,\lambda}\right)^{2/3}}}
\label{tsaddle}
\eeq
there is a saddle point in the $\theta = \pi$ direction.  This temperature
is $T_{\rm saddle} = 114.9$ MeV when the pion mass has its physical value.
Above this temperature there is no solution to the cubic equation
at $\theta = 0$.  Rather, the saddle point splits into two and circles
back around towards the origin with increasing temperature.  At a temperature
$T_{\rm ch} = 130.9$ MeV it reaches the origin.  Between $T_{\rm saddle}$
and $T_{\rm ch}$ we have a whole range of backward angles without a
minimum in the radial direction.  Above the chiral critical temperature
we have once again minima in all directions, but they are all centered
away from the origin, corresponding to $\theta \leq \pi$.
Chiral symmetry is approximately restored in the sense that
$v_{\theta}(T)$ is very small compared to $f_{\pi}$ and $T$.

The behavior depicted in figure 5 must be dependant on the use of
the linear $\sigma$--model to some extent.  Nevertheless it is
interesting to see that there is a range of temperatures where the
effective potential does not have minima in the radial direction
for parts of the backward chiral plane.  This would lead to a strong
dynamical enhancement of the forward direction in the chiral
angle distribution.  We consider this a typical example of the general
statement that the inclusion of both dynamical and thermal effects
on the evolution of distributions tends to smooth out transitions
between states and, in this sense, suppresses large angle contributions.

So far we have only discussed the volume term of the chiral angle
effective potential.  Generally one expands the free energy of a
finite system in terms of volume, surface, curvature and logarithmic
contributions.  The isotherms of figure 5 are valleys with only
one minimum located at $\theta = 0$.  Therefore no surface contribution
can be defined in principle.  Only if we include a higher order term
in $\sigma$ in the symmetry breaking potential, like a quadratic one,
can we generate a metastable configuration and therefore define a surface
free energy.  A quadratic term, for example, would generate a
barrier at $\theta = \pm \pi/2$ and in this way assure
the metastability of the condensate at $\theta=\pi$ \cite{proxi}.

To construct a surface free energy requires two phases
with equal pressures.  There are, in fact, two phases under consideration,
the quark--gluon and the hadronic phases.  Between these there is
a surface free energy (always accepting a first order phase transition).
However, the $\theta$ dependence of it is not calculable within the
linear $\sigma$-- model, nor within any model which does not incorporate
the quark and gluon degrees of freedom.  We take the point of view that
the plasma does not care in which direction the chiral field points.
The $\theta$ dependence of the surface free energy is taken to be
zero in this paper.

\subsection{Domain size distributions}

Consider a bubble which was nucleated at a temperature $T$ with
a volume $V$.  The probability that this bubble was nucleated with
a particular chiral angle $\theta$ (constant throughout its interior) is
\begin{equation}
\frac{d{\cal P}}{d\theta} \; \propto \; {\rm exp} \left[-\Delta
F_{\rm chiral}(V,T,\theta)/T \right] \, ,
\end{equation}
where $F_{\rm chiral}(V,T,\theta)/V = V_{\rm eff}(T,v_{\theta}(T),
\theta) - V_{\rm eff}(T,v_0(T),0)$.  If it is assumed that the direction
of the chiral condensate does not change from the time the bubble
was nucleated to the end of the phase transition then the joint
distribution in chiral angle and domain volume at the end of the
transition can be computed.  From eqs. (13) and (23):
\begin{eqnarray}
\frac{d^2n}{dVd\theta} = \int_{t_{\rm c}}^{t_{\rm f}} \frac{dt\,t}{t_{\rm f}}
I(t) [1-h(t)] \delta\left( V - V_{\rm bub}(t,t_{\rm f}) \right)
\frac{d{\cal P}}{d\theta} \left(V_{\rm bub}(t,t), T(t), \theta \right)
\, .
\label{above}
\end{eqnarray}
On the other hand the direction of the chiral condensate may relax
according to the probability distribution appropriate to the current
size and temperature of the bubble.  If this is the case then the
joint distribution at the end of the phase transition would be
\begin{eqnarray}
\frac{d^2n}{dVd\theta} &=& \int_{t_{\rm c}}^{t_{\rm f}} \frac{dt\,t}{t_{\rm f}}
I(t) [1-h(t)] \delta\left( V - V_{\rm bub}(t,t_{\rm f}) \right)
\frac{d{\cal P}}{d\theta} \left(V_{\rm bub}(t,t_{\rm f}), T(t_{\rm f}),
\theta \right) \,\nonumber \\
&=& \frac{d{\cal P}}{d\theta} \left(V, T(t_{\rm f}),\theta \right)\,
\int_{t_{\rm c}}^{t_{\rm f}} \frac{dt\,t}{t_{\rm f}}
I(t) [1-h(t)] \delta\left( V - V_{\rm bub}(t,t_{\rm f}) \right)
\nonumber \\
&=& \frac{d{\cal P}}{d\theta} \left(V, T(t_{\rm f}),\theta \right)\,
\frac{dn}{dV}(V)
\end{eqnarray}
where $dn/dV$ is from eq. (13).
It is difficult to know which limit more closely approximates reality.
The question is: What is the relaxation time for the chiral condensate
within a bubble?  We make an estimate of the critical damping constant
which separates the limits of weak and strong damping, analogous
to a damped harmonic oscillator, in the appendix.  For the purpose of
this paper we compute both limits which will provide bounds on what
can happen when relaxation effects are included.

To get a feel for the difference between the extreme limits of
underdamping and overdamping we plot, in figure 6, the effective
potential for the chiral angle divided by the temperature versus angle.
The Boltzmann/Gibbs factor is obtained by multiplying by a volume
and exponentiating.  The solid curve is for the temperature
at the bottom of the supercooling curve where the nucleation rate is
the highest.  The dashed curve is for the final temperature at
the end of the phase transition.  From these curves one can already
predict that the final distribution in chiral angle will be much
broader if there is overdamping; that is, if the chiral angle
can relax fast enough to follow the temperature as it goes up.

The full double differential distributions in chiral angle and
domain size is shown in figure 7.  Panels (a) and (b) represent
parameter sets A and B, respectively, with the chiral angle determined
at the moment the bubble was nucleated.
The ``angular distribution" is strongly peaked in the forward direction,
falling by one to two orders of magnitude as $\theta$ increases from
0 to $\pi/2$.
This is because the dominating nucleation temperature is around
$0.80 T_{\rm c}$ which is comparable to the chiral symmetry breaking scales
of $\sqrt{2}f_{\pi}$ (spontaneous) and $m_{\pi}$ (dynamical).
Panels (c) and (d) represent parameter sets A and B, respectively,
with the chiral angle determined at the end of the phase transition.
Since the temperature then is greater, about $0.99T_{\rm c}$, the
angular distribution is much flatter for domain sizes less than
about $30 \, V_{\rm min}$.  Thus a large relaxation rate is beneficial
to the formation of DCC at this point in the heavy ion collision.

\section{Conclusion}

We discussed the dynamical formation of disoriented chiral condensates
in ultrarelativistic heavy ion collisions. Our basic idea is that a
first order confinement phase transition governs the dynamical
evolution of the system. Once a hot plasma region is formed it will
cool according to Bjorken hydrodynamics. Bubbles or fluctuations of
the low temperature hadronic phase can form and grow after the system
supercools below the critical temperature.  This process can be
described by homogeneous nucleation theory.  Every bubble or
fluctuation can have its own chiral condensate.
The orientation of a condensate is determined
by the Boltzmann/Gibbs weight of the finite temperature effective potential
of the linear sigma model. We evaluate domain size and chiral angle
distributions.

This model is the first attempt to use the dynamics of
a first order phase transition to generate disoriented chiral
condensates. Many of the approximations we made in this model can
be relaxed in more elaborate calculations. For example, the
influence of the chiral contribution to the free energy is treated
perturbatively to lowest order. The dependance on rapidity, the
eccentricity of the domains, is neglected. The nucleation description
we apply is insufficient both at the very early stages of the
supercooling process, where very large bubbles containing several
different domains might form, as well as in the final stages of the
phase transition, where surface and/or topological effects between
overlapping domains might become important. An improved description of
the thermodynamic potential for the chiral degrees of freedom as well
as an investigation of the importance of dissipation in the  evolution
of a chiral domain embedded in a hadronic heat bath is desirable.

It is currently accepted that both the confining character of QCD as
well as the properties imposed by an approximate chiral symmetry are
crucial in determining the properties of the QCD phase transition. In
the model presented we intended to incorporate both properties in a
sensible description of the phase transition dynamics and the resulting
collective observables. We are of the opinion that modifications and
improvements of our model as well as different approaches in the same
class of models will be crucial in not only addressing issues like the
formation of DCCs but also in addressing all issues related to the
late stages of the quark--hadron phase transition.

\section*{Acknowledgements}

We thank S. Gavin for an interesting discussion during Quark Matter
'95.  This work was supported by the U.S. Department of Energy under
grant number DE-FG02-87ER40328. A.V. would like to thank the Niels
Bohr Institute for kind hospitality and financial support.

\newpage

\section*{Appendix}

In this appendix we study small amplitude oscillations of the
chiral angle about the absolute minimum of the effective potential
to a get an estimate of the time scales involved relative to the
time evolution of the temperature.  Replacing the potential in
the Lagrangian (\ref{lagrangian}) with the effective potential
in (\ref{veff}) results in the equation of motion for $\theta$.
\beq
v^2 \partial^2 \theta - Hv\sin{\theta} = 0\,.
\label{eqmot}
\eeq
This is the same equation one would have obtained by neglecting the thermal
one--loop corrections.  However, it does not include thermal effects
on the derivative terms.  The thermal effects
are incorporated via the radial field $v$ which will be constrained
to satisfy $\partial V_{\rm eff} /\partial v = 0$ in equation (\ref{vcl}).
Hence we replace $v$ in eq. (\ref{eqmot}) with $v_0(T)$ thus
eliminating its thermal fluctuations.

We would like to solve eq. (\ref{eqmot}) deep inside an
hadronic bubble where we can neglect any spatial gradients.
The equation of motion conserves energy and therefore
cannot describe the decay of $\theta$ into the absolute miniumum,
which must happen if the temperature decreases to zero.
To obtain a simple order of magnitude estimate of the time scale
in the decay process we introduce a dissipative term of the form
$\eta v^2 \partial \theta/\partial t$.  In this ansatz
$\eta$ has the dimension of energy and corresponds to a Langevin
type of friction term.  For small oscillations around zero,
relevant at low temperatures:
\beq
v_0^2(T) \, \ddot{\theta} + \eta\,v_0^2(T)
\, \dot{\theta} + H\,v_0(T) \,\theta = 0\,.
\label{teq}
\eeq
The oscillation frequencies $\omega_{\pm}$ are
\beq
\omega_{\pm} = \frac{i}{2} \, \eta \pm \sqrt{-\frac{1}{4} \,\eta^2
+ \frac{H}{v_0(T)}}\,.
\label{oszit}
\eeq
The oscillator is critically damped for the value
\beq
\eta_{\rm c}=2\,\sqrt{\frac{H}{v_0(T)}} \rightarrow 2m_{\pi} \, ,
\,\,\,\,\,\,\,\, T \ll T_{\rm ch} \, .
\label{etac1}
\eeq
Since this is a typical hadronic scale it is not clear whether
the chiral angle will be under-- or over--damped.
A stronger damping than this allows the system to relax very rapidly
on time scales of less than $\eta_{\rm c}^{-1}$.

For temperatures larger than $T_{\rm ch}$ we see from figure 5 that
the isotherms are approximately circles
in the forward half--plane with radius $1/2 v_0(T)$ centered
at ($1/2 v_0(T), 0$).  Expanding eq. (\ref{eqmot})
around the center of this circle we retain eqs. (\ref{oszit})
and (\ref{etac1}) but with $v_0$ and $\eta^{-1}$ scaled by 1/2.
Using the results from section 2 about the high temperature
behavior we find that
\begin{equation}
\eta_{\rm c} = \sqrt{\lambda} T \, , \,\,\,\,\,\,\,\, T \gg  T_{\rm ch}\,.
\label{tauc1}
\end{equation}
The friction constant would have to increase linearly with the temperature
at high temperatures to remain over--damped.  This result can be
understood as a consequence of approximate chiral restoration at high
temperatures forcing $v_0(T)$ to become small.  Actually, our small
angle approximation becomes ill--defined then.

\newpage

\vspace*{1.in}

\begin{center}
\begin{tabular}
{|c|c|c|c|c|}\hline
\,& Set A & Set B \\ \hline
$g_{\rm h}$ & $4.6$ & $7.5$ \\ \hline
$g_{\rm q}$ & $37$ & $47.5$ \\ \hline
$B^{\frac{1}{4}}$ (MeV) & $220$ & $232$ \\ \hline
$t_{\rm sc}$ (fm/c) & $6.7$ & $7.2$ \\ \hline
$T_{\rm sc}/T_{\rm c}$ & $0.81$ & $0.79$ \\ \hline
$t_{\rm f}$ (fm/c) & $30.9$ & $25.4$ \\ \hline
$T_{\rm f}/T_{\rm c}$ & $0.991$ & $0.987$ \\ \hline
$t_{\rm min}$ (fm/c) & $8.7$ & $9.7$ \\ \hline
$V_{\rm min}$ (fm$^3$)& $21.8$ & $10.1$ \\ \hline
$a$ & $0.67$ & $0.41$ \\ \hline
\end{tabular}
\end{center}
{\baselineskip=17pt
\noindent {\bf Table 1.} Summary of observables and scales for the two
different parameter sets.}
\vskip 0.5truecm
\baselineskip=20pt

\newpage

\section*{Figure Captions}

\noindent {\bf Figure 1.} The time evolution of the temperature.\\
\\
\noindent {\bf Figure 2.} The bubble volume at the end of the phase
transition as a function of the time it was nucleated.\\
\\
\noindent {\bf Figure 3.} The domain size distribution at the end of
the phase transition.\\
\\
\noindent {\bf Figure 4.} The thermal average value of the $\sigma$--field.\\
\\
\noindent {\bf Figure 5.} Polar plot of the radial minimum of the finite
temperature effective potential in the $\pi_3$--$\sigma$ plane.\\
\\
\noindent {\bf Figure 6.} The cost in free energy per unit volume
divided by the temperature as a function of the chiral angle.
The solid curve is evaluated at the moment the bubble is nucleated
while the dashed curve is evaluated at the end of the phase transition.\\
\\
\noindent {\bf Figure 7.} The double distribution
$V_{\rm min}^2 \, d^2n/dV\,d\theta$ in chiral domain size and direction
at the end of the phase transition.  In panels (a) and (b) the direction
is fixed at the moment of bubble nucleation and in panels (c) and (d)
the direction is fixed by the temperature at the end of the phase
transition.\\

\end{document}